\documentclass[pra,twocolumn,superscriptaddress]{revtex4-2}

\usepackage{nicefrac,amsmath,amssymb,mathrsfs}
\usepackage{xcolor}
\usepackage{graphicx}
\usepackage{natbib}
\usepackage{bm}
\usepackage{xspace}
\newcommand{\ofi}[0]{\textsc{of2}i\xspace}

\begin{document}

\title{Optofluidic Force Induction as a Process Analytical Technology}

\author{Marko \v{S}imi{\'c}}
\email[Corresponding author: ]{marko.simic@uni-graz.at}
\affiliation{Brave Analytics GmbH, Austria}
\affiliation{Gottfried Schatz Research Center, Division of Medical Physics and Biophysics, Medical University of Graz, Neue Stiftingtalstra\ss e 2, 8010 Graz, Austria}
\affiliation{Institute of Physics, University of Graz, Universit\"atsplatz 5, 8010 Graz, Austria}

\author{Christian Neuper}
\affiliation{Brave Analytics GmbH, Austria}
\affiliation{Graz Centre for Electron Microscopy, Steyrergasse 17, 8010 Graz, Austria}

\author{Ulrich Hohenester}
\affiliation{Institute of Physics, University of Graz, Universit\"atsplatz 5, 8010 Graz, Austria}

\author{Christian Hill}
\affiliation{Brave Analytics GmbH, Austria}
\affiliation{Gottfried Schatz Research Center, Division of Medical Physics and Biophysics, Medical University of Graz, Neue Stiftingtalstra\ss e 2, 8010 Graz, Austria}

\date{\today}


\begin{abstract}
Manufacturers of nanoparticle-based products rely on detailed information about critical process parameters, such as particle size and size distributions, concentration, and material composition, which directly reflect the quality of the final product. These process parameters are often obtained using offline characterization techniques that cannot provide the temporal resolution to detect dynamic changes in particle ensembles during a production process. To overcome this deficiency, we have recently introduced Optofluidic Force Induction (\ofi) for optical real-time counting with single particle sensitivity and high throughput. In this paper, we apply \ofi to highly polydisperse and multi modal particle systems, where we also monitor evolutionary processes over large time scales. For oil-in-water emulsions we detect in real time the transition between high-pressure homogenization states. For silicon carbide nanoparticles, we exploit the dynamic \ofi measurement capabilities to introduce a novel process feedback parameter based on the dissociation of particle agglomerates. Our results demonstrate that \ofi provides a versatile workbench for process feedback in a wide range of applications.
\end{abstract}
\maketitle

\section{Introduction}

Nanoparticles in dispersion have unique properties that make them useful in a wide range of applications including pharmaceutics, cosmetics, paint, food, and surface coatings~\cite{mitchell:20,gupta:22,kaiser:13,nile2020nanotechnologies}. To achieve the desired performance, manufacturers have to carefully monitor critical process parameters such as size and size distributions, concentration, material composition, and, if possible, the shape of nanoparticles. Conventional characterization techniques such as electron microscopy, dynamic light scattering~\cite{finsy:94,stetefeld:16}, and nanoparticle tracking analysis~\cite{hole:13} have been established as reliable offline techniques. However, they hardly provide the necessary temporal resolution to gain a deep understanding of process dynamics and their impact on critical quality attributes, which are drastically gaining relevance in modern production processes. Despite this, the complexity of a dispersion is often dominated by effects such as particle agglomeration and aggregation, which can lead to highly polydisperse samples, posing a major challenge for characterization techniques to achieve accurate and reproducible results~\cite{anderson:13,iso:16}.  

In past years, the majority of optical techniques has incorporated light for observation purposes only. When focusing light into a small volume, one can additionally exploit optical forces exerted through momentum transfer between light and matter~\cite{jonavs2008light}. In this way, nano- and micro-scaled objects  can be optically trapped in three dimensions using a tightly focused laser beam. This technique is well-known as the optical tweezers principle and dates back to 1970, when Arthur Ashkin performed his pioneering work~\cite{ashkin:70}, which he was awarded with the Nobel Prize in 2018. Modern optical traps allow for precise control of orientation, position and arrangement of particles over a broad size range~\cite{butaite2019indirect:19,donato2016light:16}. Optical tweezers have found application in many disciplines such as biology, medicine, and material sciences to investigate processes at the nano-scale, including protein–DNA interactions, protein folding, and molecular motors~\cite{bustamante2021optical}. Their non-invasive character makes optical traps a versatile and attractive tool for the analysis and characterization of nanoparticles. 

Building upon the concept of optical tweezers, in optical chromatography~\cite{imasaka:95} a weakly focused laser beam is used to establish a two-dimensional optical trap for particles of different size, which are pumped into a measurement cell. Along the optical axis of the incoming laser beam the optical forces counteract the fluidic ones. For appropriately chosen laser power and flow velocities, particles eventually come to a complete halt in the measurement cell at different positions, which can then be directly related to particle size. While this technique has been applied successfully for single particle analysis in the past, it has a limited capability for processing large amounts of particles with high throughput.

In a recent paper~\cite{simic:22}, a novel characterization method, Optofluidic Force Induction (\ofi), has been introduced. In contrast to optical chromatography the laser beam and fluid propagate in the same direction, such that the nanoparticles to be analyzed experience size-dependent velocity changes, which are observed and can be used to extract particle size distributions. This allows for single particle characterization online and in real-time, even for highly polydisperse and multi modal samples. To avoid collisions of particles in the focal region, we use a vortex beam with zero intensity at the optical axis. A flow-through measurement cell then enables fast processing of a large number of particles as well as measurements of ultra-low concentrated samples, where low sample availability might play an important role. The real-time capabilities of the system make it attractive for many applications, as it can reveal dynamic changes in form of process feedback parameters in response to changing conditions during a production run. 

In this paper, we apply \ofi to industry relevant samples and demonstrate its capabilities for online and real-time process feedback. In the context of nanoparticle characterization, we provide continuous information about particle count, size, and size distributions, as well as concentration at laboratory scale. Additionally, by exploiting the dynamic measurement principles of the system, a novel process feedback parameter is introduced to study particle agglomeration and dissociation processes. Further, we discuss possible directions for future extensions towards the extraction of information about forward scattered light as well as spectroscopic information to gain deeper insights into the material composition of particle ensembles.

\begin{figure}[t]
	\includegraphics[width=\columnwidth]{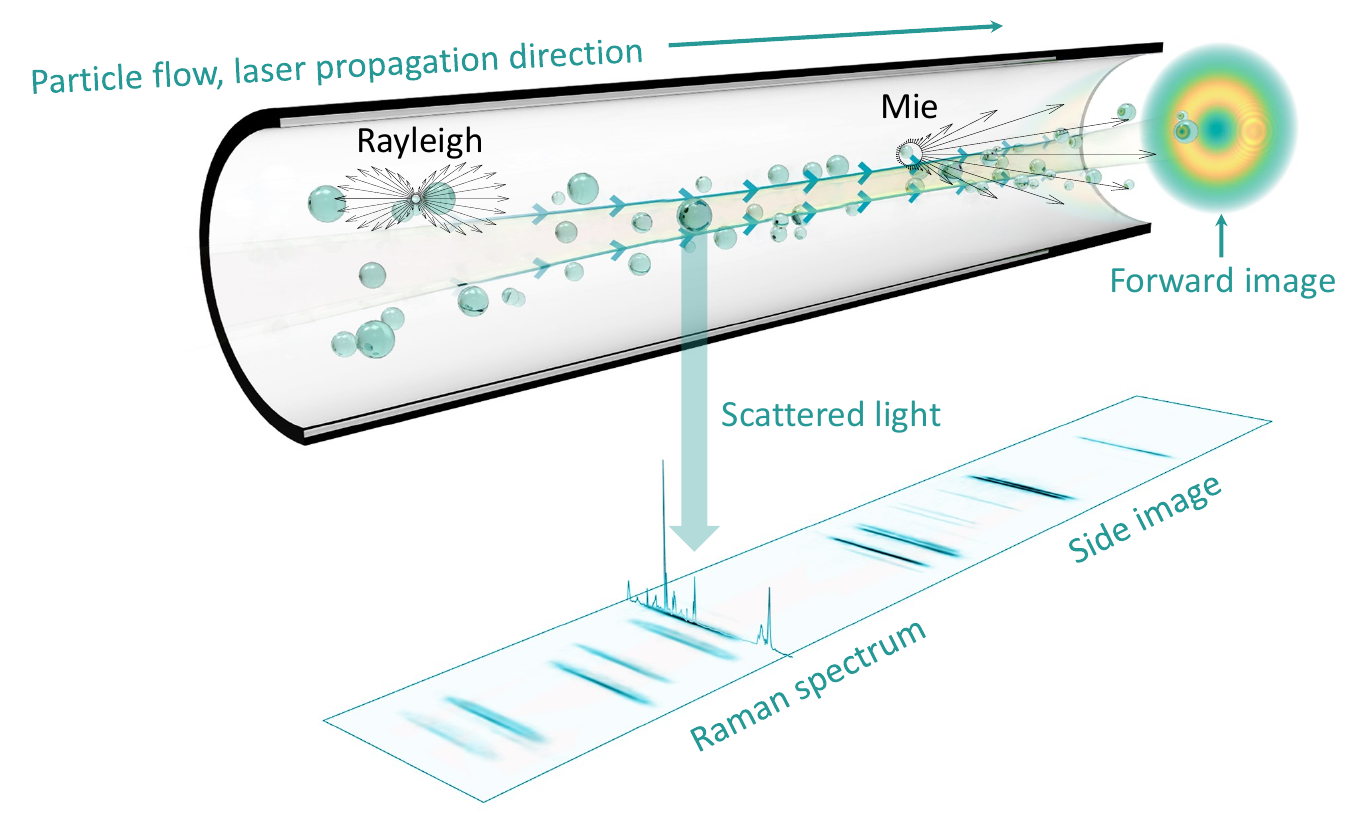}
	\caption{Schematics of optofluidic force induction (\ofi). (Top)~Nanoparticles to be analyzed are transported through a microfluidic channel alongside a weakly focused laser beam with an optical vortex (optical angular momentum $m=2$).  The purpose of the laser is threefold.  First, through the optical forces in the radial directions nanoparticles sufficiently close to the intensity maxima become trapped  in the transverse directions.  Second, the optical forces in the laser propagation direction push the particles and lead to velocity changes depending on size and material properties.  Third, the light scattered off the particles is detected and allows monitoring the velocity changes. (Bottom) Scattered light intensity as recorded by a camera. Each particle appears as a line in the side image due to astigmatism caused by the capillary. Scattered light of large particles is recorded in the forward image as a diffraction pattern.}
	\label{fig:OF2iSketch}
\end{figure}
\section{Materials and methods}\label{sec:matsmeth}
\subsection{Materials}\label{sec:materials}
\noindent In our experiments we used three different types of samples:
Polystyrene (PS) spheres (Thermo Scientific, 3000 Series NIST\textsuperscript{\texttrademark} traceable  Nanosphere\textsuperscript{\texttrademark} Size Standards, $\approx\!1\%$ polystyrene solids in 15 ml aqueous suspension) with nominal diameters of 203 nm ($\pm 4$ nm), 401 nm ($\pm 6$ nm), 600 nm ($\pm 9$ nm), and PS spheres (Applied Microspheres, 20000 series NIST\textsuperscript{\texttrademark} traceable particle size standard and count control, $\approx\!1\%$ polystyrene solids, $< 0,1$\% surfactants and $< 0,05$\% preservatives in 20 ml aqueous suspension) with nominal diameters of 789 nm ($\pm 22$ nm) and 1040 nm ($\pm 28$ nm) were purchased commercially. For the preparation of the Ni-P/SiC electrolyte solution following chemicals were provided by our industrial project partner: Nickel sulfate hexahydrate ($\text{NiSO}_{\text{4}}\text{6H}_{\text{2}}\text{O}$, NSH), nickel chloride hexahydrate ($\text{NiCl}_{\text{2}}\text{6H}_{\text{2}}\text{O}$, NCH), phosphoric acid ($\text{H}_{\text{3}}\text{PO}_{\text{4}}$, PA1), phosphorous acid ($\text{H}_{\text{3}}\text{PO}_{\text{3}}$, PA2), saccharin sodium salt ($\text{C}_{\text{7}}\text{H}_{\text{4}}\text{NO}_{\text{3}}\text{SNa}$), sodium lauryl sulfate ($\text{C}_{\text{12}}\text{H}_{\text{25}}\text{Na}\text{O}_{\text{4}}\text{S}$) (SDS), sodium hydroxide ($\text{NaOH}$), sulfuric acid ($\text{H}_{\text{2}}\text{SO}_{\text{4}}$) and silicon carbide (SiC) nanoparticles with nominal diameter of 100 nm.
The oil-in-water emulsions were taken from the H1 and H2 stage of a high-pressure homogenization process, respectively, where a H1 sample is obtained by further homogenizing a H2 sample.

\subsection{Preparation of Polystyrene mixture}\label{sec:preppolystyrene}
For the PS dispersion, a mixture of 203 (200~$\mu\rm L$), 600 (200~$\mu\rm L$), and 1040~nm (300~$\mu\rm L$) PS spheres was prepared initially. During the measurement, 401~nm (200~$\mu\rm L$) PS spheres were titrated into the initial mixture. Similarly, about ten minutes after the 401~nm particles were titrated, 789~nm (300~$\mu\rm L$) PS spheres were titrated into the mixture. The dispersion was prepared in a way that the total number of particles detected by \ofi does not vary greatly during the monitoring process. For the dilution factors see Sec~\ref{sec:sampledilution}.

\subsection{Preparation of Ni-P/SiC electrolyte solution}\label{sec:prepsic}
First, an electrolyte solution was prepared by dissolving NSH (260 g/L), NCH (48 g/L), PA1 (40 g/L) and PA2 (20 g/L) into deionized water (50\% of total water volume). The pH was adjusted to a value of 1.5 through the addition of saturated sodium hydroxide solution (1.25 M) and sodium saccharin salt (2 g/L). In a separate step, SDS (2.5 g/L) was dissolved in deionized water (10-20\% of total water volume). The SiC nanoparticle powder (10 g/L) was added to the SDS solution and remained under constant stirring for 30 minutes to reduce particle agglomeration. In the last step, the SDS/SiC nanoparticle mix was added to the electrolyte solution. The Ni-P/SiC solution temperature was kept constant at 50$^{\circ}$C using a hot plate stirrer. Again, saturated sodium hydroxide and water were added to maintain a pH value of $2.00\pm0.02$. 

\subsection{Sample dilution}\label{sec:sampledilution}
\noindent To comply with our experimental setup, we diluted highly concentrated samples in ultrapure water (MilliQ\textsuperscript{\textregistered} type 1), 0.02 $\mu\text{m}$ microfiltered through an inorganic sterile membrane filter (Whatman Anotop\textsuperscript{\texttrademark}25; Diameter: 25 mm; Pore size: 0.02 $\mu$m). Dilution factors for the pure PS samples are 1:40\,000 (203 nm), 1:10\,000 (401 nm), 1:5\,000 (600 nm). The dilution factor for the oil-in-water emulsions is 1:200\,000 (H1, H2). The Ni-P/SiC electrolyte solution was diluted using our in-house developed dilution system with a dilution factor of 1:2\,000 in continuous mode.

\begin{figure*}[t]
	\includegraphics[width=0.7\textwidth]{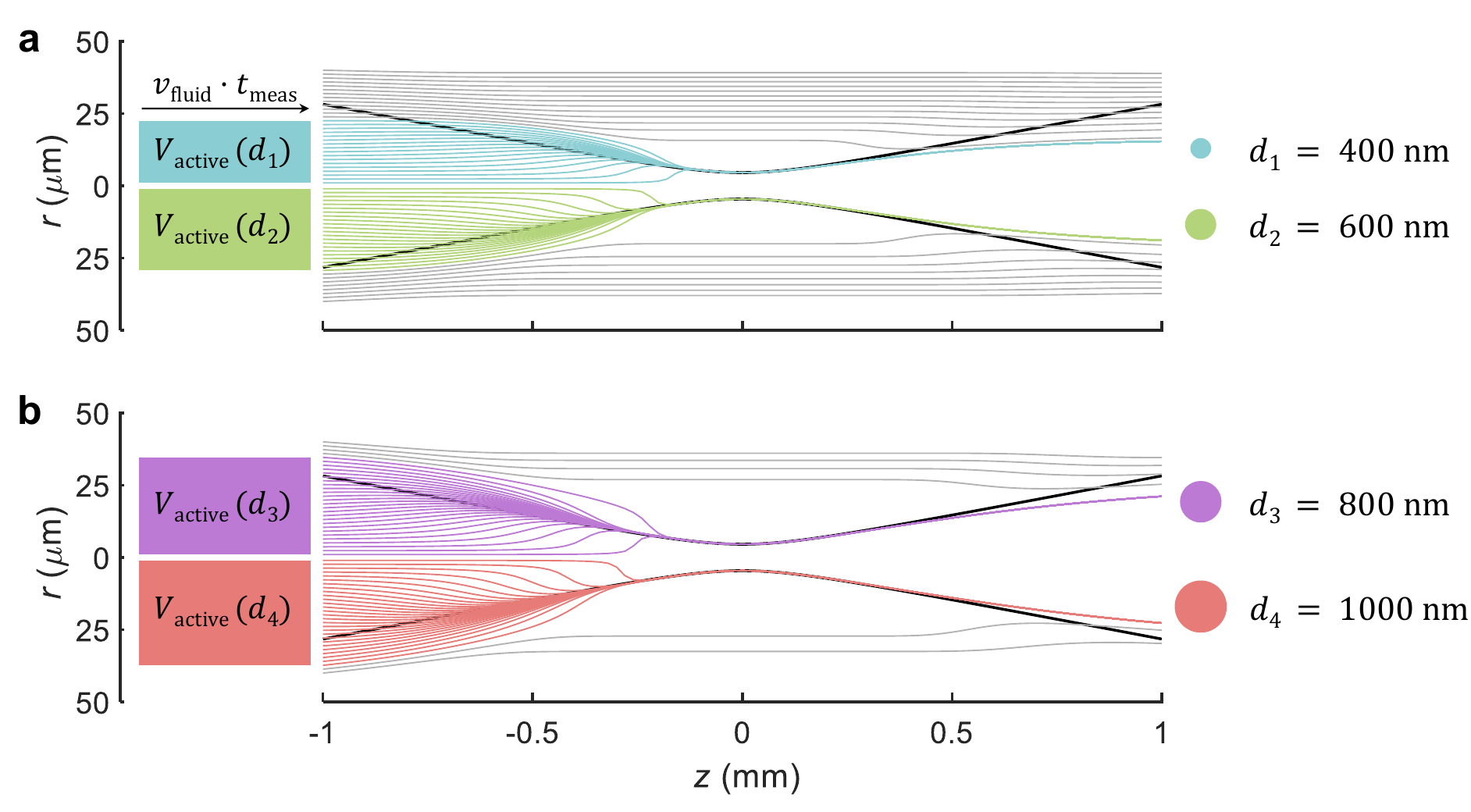}
	\caption{2D projection of the active volume and simulated trajectories for different transverse starting positions $r$ and diameters of polystyrene nanoparticles ($n_p = 1.59$). The initial positions $(r,0,z_0)$, laser, and fluid parameters are the same for all simulations, with $z_0=-1~\text{mm}$ chosen sufficiently far away from the focal region where the optical forces are negligible. Particles become either trapped (colored trajectories) or not (gray trajectories), where the active volume and trajectories of trapped particles are color coded with respect to their size. With increasing sphere diameter, more particles are trapped as the optical forces increase for the larger particles. This in turn leads to an increase of the active volume. Note that for better visualization in the above panels we only plot the active volume above or below the optical axis.}\label{fig:activevolume}
\end{figure*}

\subsection{OF2i principle}\label{sec:of2i}
The basic principle underlying \ofi is shown in Fig.~\ref{fig:OF2iSketch}: nanoparticles under investigation are dispersed in a solution and are pumped through a microfluidic measurement cell. In addition, a weakly focused vortex beam (topological charge $m = 2$) propagates in the direction of the flow. Particles sufficiently close to the laser beam become optically trapped in the radial direction (gradient forces) and experience velocity changes along the propagation direction $z$ of the laser beam (scattering forces). The light scattered off the individual particles is monitored outside the measurement cell. Note that the capillary acts as a cylindrical lens, thus each particle is imaged as a line~\cite{meinert:17} (see side image in Fig.~\ref{fig:OF2iSketch}). By analyzing the velocity changes as particles travel through the focal region of the beam, one obtains information about their sizes. In the future, we plan to extract further information from the forward scattered light and the emission pattern, particularly for large particles, as well as from Raman spectra.

\subsection{Theory}\label{sec:theory}
Our theoretical framework is based on Maxwell's equations and Mie theory~\cite{simic:22,simic:23}. We have developed a four-step model for the simulation of nanoparticle trajectories in the flow cell in presence of a weakly focused vortex beam, which leads to trapping and velocity changes of the nanoparticles. The first three steps are concerned with Maxwell's equations where we (i) provide an expression for the incoming Laguerre-Gaussian laser beam, (ii) solve Maxwell's equations in presence of a nanoparticle, and (iii) use the total fields (sum of incoming and scattered fields) to compute the optical force $\bm F_{\rm opt}(\bm r)$ acting on a nanoparticle at position $\bm r$. In step (iv), we compute the trajectory from Newton's equation of motion,
\begin{equation}\label{eq:newton}
	m\ddot{\bm r}=\bm F_{\rm opt}(\bm r)+\bm F_{\rm drag}+\bm F_{\rm brownian}\,,
\end{equation}
where $m$ is the mass of the nanoparticle, which might include the added mass due to the fluid, $\bm F_{\rm drag}$ is the drag force acting on the particle moving through the fluid, and $\bm F_{\rm brownian}$ accounts for the stochastic forces due to thermal fluctuations also known as Brownian motion~\cite{kubo:85,einstein:05}. For the force acting on a sphere moving through a viscous fluid with velocity $\bm v$ we consider Stoke's drag, 
\begin{equation}
	\bm F_{\rm drag}=-6\pi\mu R\big(\bm v-\bm v_{\rm fluid}\big)\,,
\end{equation}
where $\bm v_{\rm fluid}$ and $\mu$ are the velocity and dynamic viscosity of the fluid, respectively, and $R$ is the radius of the sphere. For sufficiently large spheres, say for diameters above 10 nm, the momentum relaxation time is so short that we can approximately set $\dot{\bm v}\approx 0$~\cite{neuman:08}. Also, the Brownian motion doesn't play a decisive role for larger sphere, as previously discussed in~\cite{simic:23}. For the condition that the optical force is balanced by the drag force, we get for the particle velocity
\begin{equation}\label{eq:vsteady}
	\bm v(\bm r)=\bm v_{\rm fluid}+\frac{{\bm F}_{\rm opt}(\bm r)}{6\pi\eta R}\,.
\end{equation}
When inferring the particle number distribution from \textsc{of}2i measurements, we have to account for the fact that larger particles become trapped more easily than smaller ones, owing to the increase of optical forces with increasing particle size. 

Fig.~\ref{fig:activevolume} reports simulated trajectories for particles with diameters of 400, 600, 800 and 1000 nm and refractive index of $n_p = 1.59$ (PS). The flow velocity $v_{\rm{fluid}} = 0.1~\rm{mm/s}$, the refractive index of the flow medium $n_b = 1.33$ (water), and the incoming beam power $P = 1.65~ \rm W$ are the same for all simulations. The particles start at $(r,0,z_0)$, where $z_0 = -1~\rm{mm}$ is chosen sufficiently far away from the focus region, such that the optical forces are negligible, and then propagate in presence of optical and fluidic forces along $z$. We observe that particles in the focus region are either trapped or not. Thus, we define a cutoff parameter $r_{\rm{cut}}(d,n_p)$ as a function of particle diameter $d$ and refractive index $n_p$ for which particles with a given transverse starting position $r \leq r_{\rm{cut}}(d,n_p)$ are trapped and accelerated towards the focus region of the beam. As previously discussed in~\cite{simic:22}, we can define an active volume 
\begin{equation}\label{eq:vactive}
	V_{\text{active}}(d,n_p) = \left[\pi r_{\text{cut}}^2(d,n_p)\right] v_{\text{fluid}}\, t_{\text{meas}}\,,
\end{equation}
where the term in brackets is the cross section in transverse direction and $v_{\rm{fluid}}\,t_{\text{meas}}$ is the sampling distance spanned along the propagation direction of the flow. 

For a detailed description of our simulation approach, we refer the reader to~\cite{simic:23}, where we have covered the full working equations and model ingredients, influence of Brownian motion and refractive index. Further, we emphasize that our model contains no free parameters and all laser, fluid and nanoparticle parameters can be inferred from experiment.

\subsection{Experimental setup}
The experimental setup of \ofi builds on a two-dimensional optical trap in a microfluidic flow channel of cylindrical shape~\cite{simic:22}, as exemplary shown in Fig.~\ref{fig:OF2iSketch}. A linearly polarized laser beam is generated by a 532 nm CW DPSS laser (Laser Quantum, GEM532). The beam alignment is performed using two mirrors and a 5x beam expander. An azimuthally polarized Laguerre-Gaussian laser mode with topological charge $m=2$ is generated using a zero-order vortex half-wave retarder ($q=1$). A converging lens is used to focus the vortex beam into a measurement cell. A microfluidic pump is connected to a multi-port valve and transports the sample to be analyzed into the measurement cell where the light-matter interaction takes place (see Fig.~\ref{fig:OF2iSketch}). The scattered light is magnified by an ultramicroscope setup consisting of a 10x PLAN microscope objective, an optical filtering bank, a 75 mm focusing lens, and a CMOS camera providing the raw video signal (see side image in Fig.~\ref{fig:OF2iSketch}). A second imaging path is used to monitor the cross-section of the illuminating beam and to observe forward scattered light. Particles leaving the focal region are transported to waste via an outlet port. For schematics of the OF2i setup, we refer the reader to~\cite[Fig. 7]{simic:22}. A computer running BRAVE Analytics proprietary software suite HANS provides the user with control of laser output power, flow direction and speed, and valve positions via a user interface. Additionally, live visualization of both imaging systems is provided by the software. The above stated parameter set and experimental setup were chosen to fit the broad range of applications reported in this paper, with a dynamic size range of about 150 nm to 5 $\mu$m. To deal with smaller or larger particles the optical setup might have to be adjusted accordingly.

\subsection{Data processing}\label{sec:dataprocessing}
\noindent Raw video data was recorded at 200 fps using HANS software suite 2.3 and processed via  \textsc{matlab} routines, covering the (i) generation of waterfall diagrams and trajectories from single particle light scattering, (ii) velocity determination via the trajectory's slope, and (iii) computation of number-based particle size distributions from time series. For the velocity determination, we start by computing a waterfall diagram from the raw input data by integrating along the vertical axis of each frame. Such a diagram, denoted $I(t,z_i)$ and shown as exemplary in Figure~\ref{fig:SiCmonitoring}(a), encodes the sideways scattered light by a particle at position $z_i$ and video frame $t$. In the next step, we extract images for given positions in the focal region $z_f$ within a range of $\pm30$ pixels. In a second image we plot a straight line with gradient angle $\theta$ weighted by a Gaussian kernel ($\sigma=2$) in transverse direction. We choose $\theta$ such that the overlap of both images is maximized. The velocity of a particle is then calculated from $\theta$. Particle position in pixel format was transformed to a local coordinate system using a calibration constant of 0.7867 $\mu \text{m}$/pixel. For cluster analysis of particle ensembles, a one-dimensional Gaussian mixture model (GMM) with five components was applied. The trend analysis was performed using the BEAST algorithm, a Bayesian model for time series decomposition and change point detection~\cite{zhao2019detecting}. 
In the following, we discuss the main steps to obtain particle size from the measured velocities.

\subsection{Particle size distribution}
Here we describe the steps performed to obtain a number-based particle size distribution (PSD) within \ofi. First, we start with the simulation of particle trajectories for particles with given size, refractive index, and system parameters (e.g. laser power, flow velocity) based on our model discussed in~\ref{sec:theory}. From the trajectories, we extract the velocity profile of particles as they travel through the capillary. For trapped particles, we extract the corresponding maximum velocities $v_{f, \rm expt}$ in the focal region $z_f$ of the laser beam, which can be compared with those of Eq.~\eqref{eq:vsteady}. The maximum velocities and cutoff parameters are used to set up a lookup table that relates the simulated particle velocity to the corresponding size (for given refractive index $n_p$), 
\begin{equation}\label{eq:vtod}
	d=d(v_{\rm expt},n_p)\,.
\end{equation}
Suppose that within a given time interval $t_{\rm meas}$ a certain number of particles $\bar{N}(d,\Delta d)$ with a diameter in the range of $(d,d+\Delta d)$ is observed. To obtain the PSD $\bar n(d)$, we have to divide this number by the active volume,
\begin{equation}\label{eq:psdact}
	\bar n(d)\Delta d=\frac{\bar{N}(d,\Delta d)}{V_{\rm active}(d,n_p)}\,.
\end{equation}
Here, $V_{\rm active}$ corrects for the fact that larger particles are trapped more easily than smaller ones, because of the size-dependent trapping forces, and are consequently observed more frequently in \ofi. See also Fig.~\ref{fig:activevolume} and the discussion above. Thus, once Eq.~\eqref{eq:vtod} and the expression for the active volume are given from our theoretical model, we can compute the PSD directly from Eq.~\eqref{eq:psdact}.

\section{Results and Discussion}\label{sec:results}
In this work, we have performed measurements of highly polydisperse and industrially relevant samples. We next present our results, which are divided into three sections: First, we start by demonstrating the capability of \ofi to measure polydisperse and multi modal samples on the example of standardized polystyrene spheres with well-known size distributions and narrow standard deviation (for details see Sec.~\ref{sec:materials}). Second we present our findings for an oil-in-water emulsion stemming from a high-pressure homogenization process. Finally, we introduce a novel process feedback parameter based on the dissociation processes of SiC nanoparticle agglomerates. All measurements have been performed with a constant beam power and flow rate of $P=1~\rm W$ and $\dot{V} = 4~\mu\rm{L/min}$, respectively, unless stated otherwise.

\section{Results and Discussion}\label{sec:results}
In this work, we have performed measurements of highly polydisperse and industrially relevant samples. We next present our results, which are divided into three sections: First, we start by demonstrating the capability of \ofi to measure polydisperse and multi modal samples on the example of standardized polystyrene spheres with well-known size distributions and narrow standard deviation (for details see Sec.~\ref{sec:materials}). Second we present our findings for an oil-in-water emulsion stemming from a high-pressure homogenization process. Finally, we introduce a novel process feedback parameter based on the dissociation processes of SiC nanoparticle agglomerates. All measurements have been performed with a constant beam power and flow rate of $P=1~\rm W$ and $\dot{V} = 4~\mu\rm{L/min}$, respectively, unless stated otherwise.

\subsection{Monitoring of polydisperse samples}\label{sec:monitoringPS}
\begin{figure*}[t]
	\includegraphics[width=0.9\textwidth]{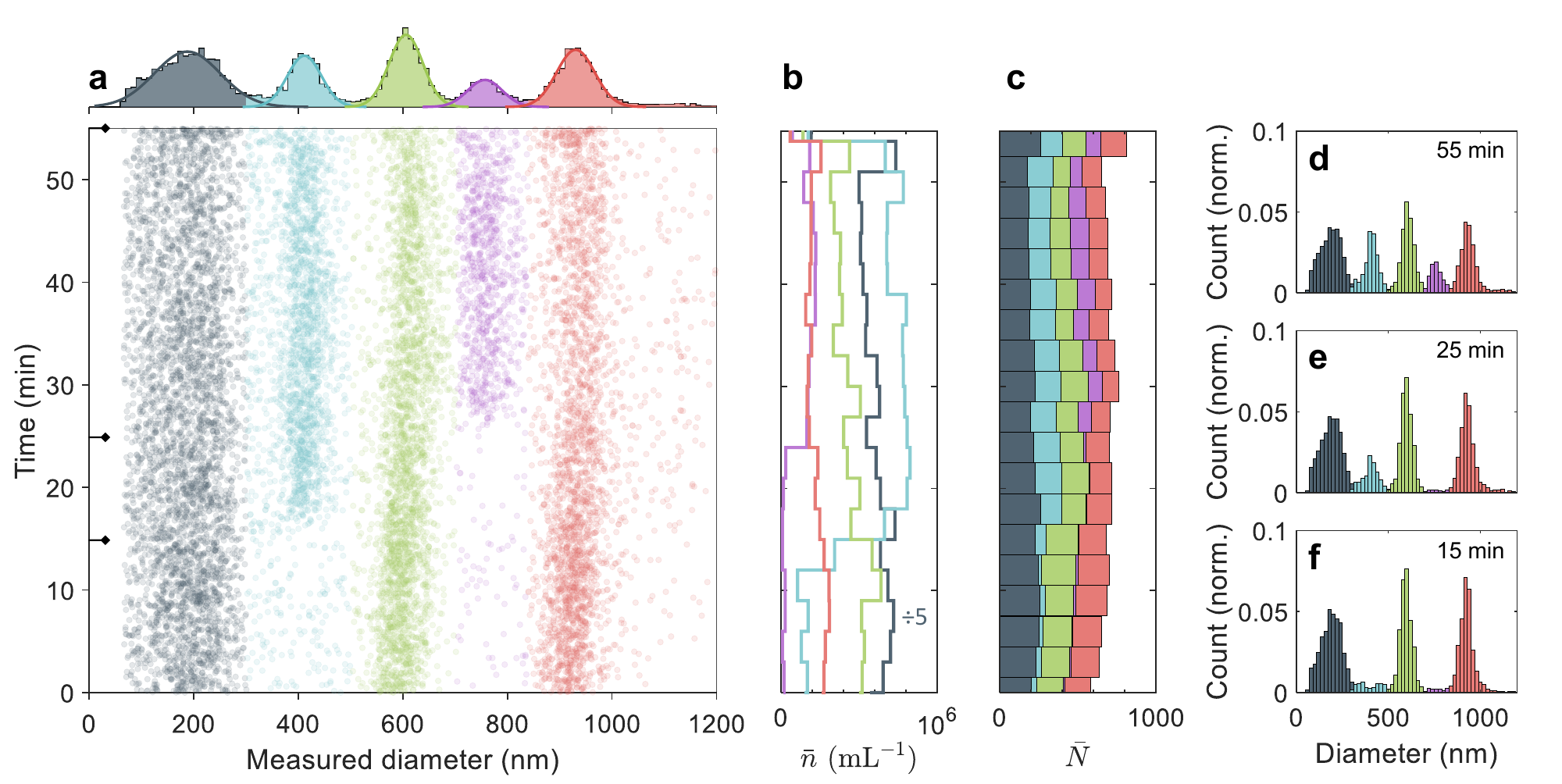}
	\caption{Continuous monitoring of a polydisperse sample by \ofi. (a) Measured diameters of a mixture of PS spheres ($n=1.59$) with nominal diameters of 203, 401, 600, 789, and 1040~nm over a large measurement time. The contribution of each ensemble is determined by a 1D Gaussian mixture model with five components applied to the histogram on top and highlighted by color. (b) Total concentration $\bar{n}$ (Particles/mL) and (c) count $\bar{N}$ of each cluster, averaged over a period of $t_{\rm{meas}} = 180~\rm s$. For the smallest particles (203 nm), the concentration is divided by a factor of five for better representation. Particle size histograms ($\Delta d = 10~\rm{nm}$) obtained after (d) 55, (e) 25, and (f) 15 minutes, indicated by markers on the y-axis in panel (a).}
	\label{fig:polystyrenemonitoring}
\end{figure*}
\noindent One of the major challenges in particle characterization is the handling of highly polydisperse and multi modal samples, which are frequently encountered in industrial processes. To demonstrate \ofi's capability in characterizing such samples, we have prepared a dispersion, initially consisting of PS spheres ($n_p = 1.59$~\cite{sultanova:09}) with nominal diameters of 203, 600, and 1040 nm immersed in water ($n_b = 1.33$), which span a broad size range. During the monitoring process, two additional ensembles with nominal diameters of 401 and 789 nm were titrated into the initial dispersion at different times with a total delay of approximately ten minutes for each ensemble, to introduce dynamic changes in the dispersion (for details see Sec.~\ref{sec:preppolystyrene}).

Fig.~\ref{fig:polystyrenemonitoring} shows results for a continuous measurement of PS spheres, as obtained by \ofi and computed according to the scheme discussed above. Let us first focus on the scatter diagram in panel (a), where each data point represents an individually measured particle. By closer inspection of the data, we observe three distinct particle populations from the very start of the measurement throughout the experiment. See also the three sharp peaks in the particle size histogram shown in panel (f), obtained after 15 minutes of measurement. After about 17 and 27 minutes, respectively, two more populations emerge in the scatter plot from the titrated PS samples (401 nm, 789 nm) and are monitored until the end of our measurement. Again, the individual peaks can be clearly identified as shown in panels (e) and (d). To identify the various contributions of each ensemble, a 1D GMM model with five components (full covariance) was applied to the total particle size histogram on top of panel (a). The corresponding contributions are highlighted by color and fitted by a Gaussian distribution (solid line). Note that some low counts in panel (f) are identified by the 1D GMM and are most likely due to particle agglomerates, see also the scatter diagram in panel (a). 

At this point, two conclusions can be drawn. First, the continuous measurement of the PS dispersion allows us to observe dynamic changes in the dispersion. Second, the weakly focused vortex beam incorporated into the system enables parallel measurement of single particles over a broad size range. This can be seen in particular in panel (a), where we are able to detect a very low concentration of particles in the size range $ d > 1000$~nm, which we attribute to agglomerates in the dispersion, see also the tailing of the PSD in panel (d). In general, we see a very good agreement between the measured PSD peaks and the nominal diameters specified by the manufacturer, except for the largest particles (1040 nm), whose size is underestimated in our system. This might be due to a slight deviation in the assumed refractive index, however, further work is needed to clarify this point. Note also a slight shift of the measured diameter which we attribute to a power shift of the laser.

Figures \ref{fig:polystyrenemonitoring}(b) and \ref{fig:polystyrenemonitoring}(c) show the total concentration $\bar{n}$ (particles/mL) and particle count $\bar{N}$, respectively, as obtained by continuous monitoring and computed by Eq.~\eqref{eq:psdact} ($\Delta d = 10~\rm{nm}$). We chose a time average of $t_{\rm{meas}} = 180~\rm s$. For this experiment, the active volume was calculated using the nominal diameters of the PS spheres entering Eq.~\eqref{eq:vactive} (see also Fig.~\ref{fig:activevolume}). In panel (b), each line represents the total concentration (this is $\bar{n} = \sum_{d}\bar{n}(d)$ within $t_{\rm{meas}}$) for the corresponding clusters identified by the GMM. Let us first focus on panel (c), where we observe a relatively constant particle count, whereas the contribution of each population varies over time. 

To maintain the total particle count throughout the experiment, the initial dilution factor of each population had to be chosen appropriately. In this case, the amount of small particles (less frequently trapped) in the initial solution is chosen to be higher compared to larger particles. This is shown by the separate lines in panel (b), where the correction by the active volume, Eq.~\eqref{eq:vactive}, becomes apparent. As expected, our results show a significantly higher concentration for the smallest particles compared to other populations. In the figure we divide the calculated concentration by a factor of five for better representation. Our results confirm that \ofi is capable of monitoring particle size, size distribution, number, and concentration of highly polydisperse and multi modal samples over a remarkable size range in real time. Furthermore, we checked that our measurement scheme produces stable and reproducible results for continuous measurement over a time period of at least one hour.

\begin{figure*}[t]
	\includegraphics[width=0.85\textwidth]{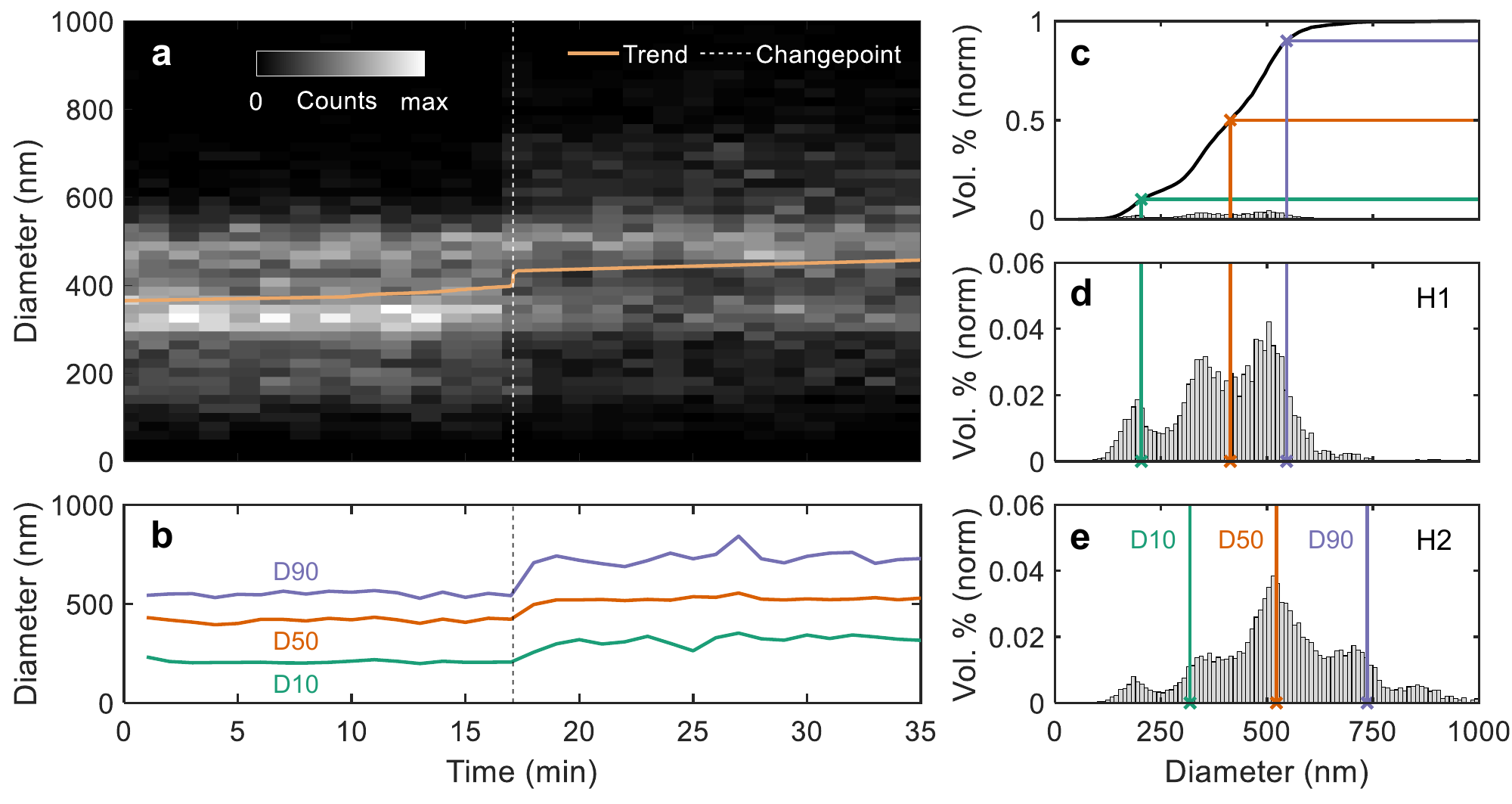}
	\caption{Continuous monitoring of oil-in-water emulsions ($n_p = 1.46)$ stemming from two stages (H1, H2) of a high pressure homogenization process. (a) Number based 2D-histogram and trend line (mean) calculated from measured particle diameters. The dashed line marks the transition from H1 to H2 at the change point $t_{\rm{cp}} = 17~\rm{min}$ as determined by the BEAST algorithm. (b) Corresponding D values as a function of time, extracted from the cumulative PSD (volume based) which is accumulated over a time period of $t_{\rm{meas}} = 60~\rm s$. D values represent the diameter below which 10\%, 50\%, and 90\% of the observed particle volume is detected. (c) Cumulative PSD ($\Delta d = 10~\rm{nm}$) obtained from the underlying distribution shown in (d). (d), (e) Normalized volume based PSD for the two stages H1 and H2, respectively. The D values are marked by crosses at the horizontal axis and show a shift towards larger particles from H1 to H2.}
	\label{fig:emulsionmonitoring}
\end{figure*}
\subsection{Monitoring of oil-in-water emulsion}
In what follows, we demonstrate the applicability of \ofi to infer detailed information about process dynamics of industrially relevant samples. In modern production processes, manufacturers often rely on offline characterization methods to check the quality and performance of final products. This can lead to long downtime and production of large amounts of waste if failure or abnormalities within the process are not detected at an early stage. Together with one of our industrial partners, we have implemented \ofi for a continuous monitoring of an established process for the production of oil-in-water emulsions. Our main goals are to gain a deeper understanding of the process dynamics and to improve the quality and consistency in the final product. In addition, we want to detect changes in the PSD as soon as possible, and keep the concentration of large particles as low as possible to reduce waste and downtime during the process.

\begin{figure*}[t]
	\includegraphics[width=\textwidth]{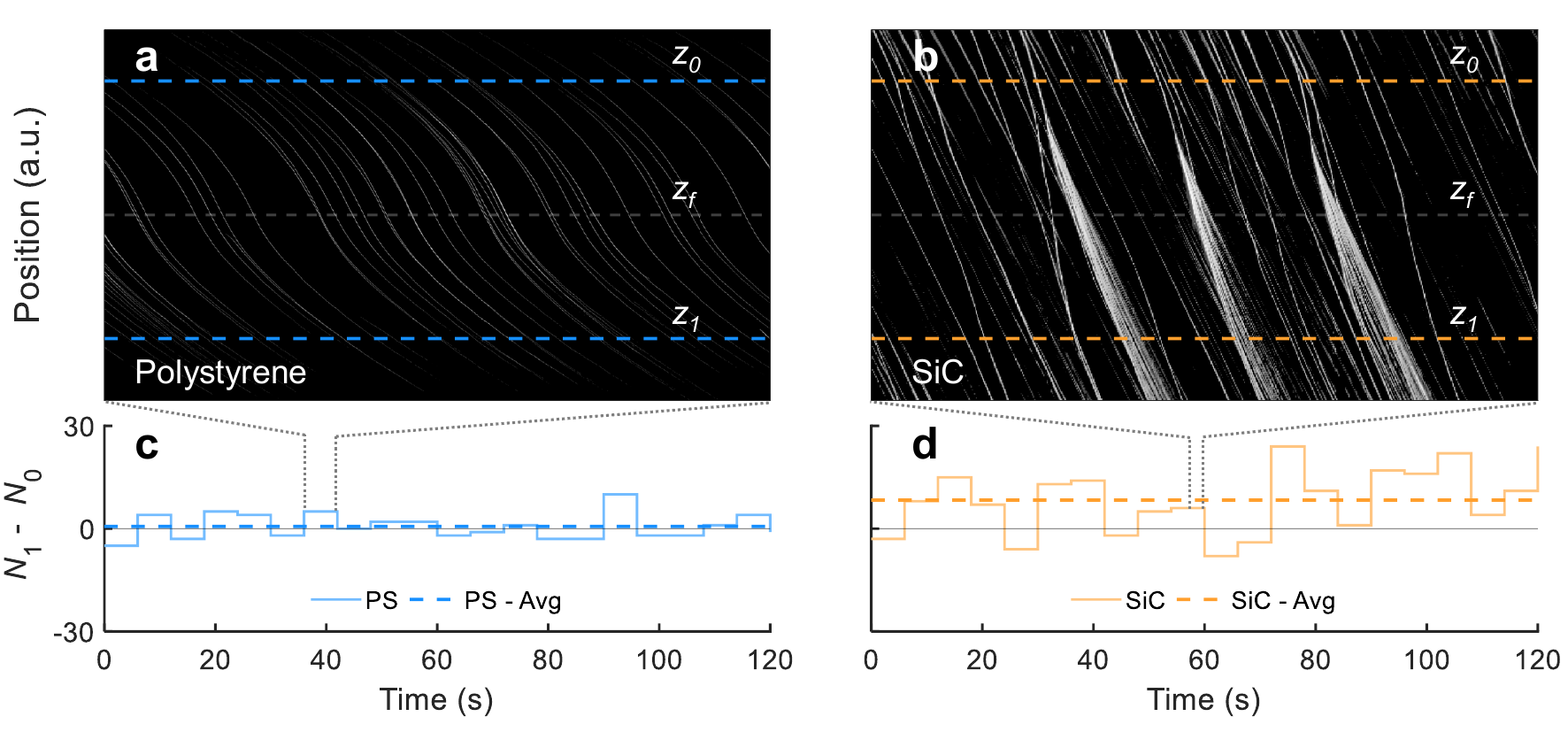}
	\caption{Results of \ofi measurements for PS nanoparticles in water and SiC nanoparticles immersed in a Ni-P electrolyte solution. The figure shows waterfall diagrams obtained from scattered light and single particle trajectories of (a) PS and (b) SiC nanoparticles, respectively, obtained from raw video data (see also Sec.~\ref{sec:dataprocessing}). For the SiC nanoparticles one observes dissociation of particle agglomerates as they approach the focal region $z_f$ of the laser beam. We perform counts $N_0$ and $N_1$, respectively, at the reference points $z_0$ and $z_1$, marked as dashed lines. The difference of the number of particles entering and leaving the measurement cell is shown in (c) and (d), respectively, as a stair plot and averaged over $t_{\rm{meas}} = 6~\rm s$. The dashed lines in (c),(d) show the average of the solid line with a significant increase above zero for the SiC nanoparticles indicating dissociation of SiC agglomerates. The dotted lines indicate the corresponding waterfall diagram. }
	\label{fig:SiCmonitoring}
\end{figure*}

To investigate the performance of \ofi on the example of a complex emulsion, we have performed measurements on two representative samples (spherical shape, $n_p = 1.46$) at laboratory scale. These were taken from different stages of a high-pressure homogenization process, which we refer to as H1 and H2, respectively, where H2 is expected to contain larger particles. Both samples were diluted to comply with our experimental setup. Initially, we started by monitoring H1 and after some time introduced H2, resulting in a total measurement duration of 65 minutes. Fig.~\ref{fig:emulsionmonitoring}(a) shows as a function of time the histogram of measured diameters, obtained by continuous monitoring of the emulsions. In contrast to the PS samples previously discussed, we observe a very broad size distribution throughout the monitoring process. We also note that after some time we start to detect larger particles in the size range $> 600~\rm{nm}$. To gain a better understanding of the process dynamics, we have performed a trend and change point analysis by applying the BEAST algorithm (mean mode) to the calculated particle diameters (see also Sec.~\ref{sec:dataprocessing}). The resulting trend (solid line) and change point (dashed line) are plotted in the same figure as overlay. The trend line reveals a slight increase after 10 minutes with an abrupt change towards larger particle diameters at $t_{\rm{cp}} = 17~\rm{min}$, followed by a continued increase. Here $t_{\rm{cp}}$ is the change point detected by the software. This does not come unexpectedly. Once the H2 sample flows through the measurement cell, larger particles are detected more frequently by~\ofi. Therefore, in our analysis, we assign particles measured before $t_{\rm cp}$ to H1 and after $t_{\rm cp}$ to H2, respectively. For this kind of samples, it is often preferred to report measurement results as a volume distribution, which gives more weight to larger particles~\cite{burgess2004particle}. Similar to the analysis of PS spheres discussed above, we begin by calculating a number-based PSD taking into account the active volume. From there we compute the volume of the individually monitored spheres within a small size range ($\Delta d = 10~\rm{nm}$) as well as the volume of the whole ensemble. 

Panels (d) and (e) report the normalized volume distribution obtained for the two states H1 and H2, respectively. When comparing the results for the two stages in the process, we observe three matching peaks in both PSDs, whereas the second peak around 350~nm is significantly higher in H1 compared to H2. The PSD in (e) features a peak around 720~nm with a long tailing up to the micro meter regime. Although both the trend analysis and PSDs provide valuable information, we seek for an alternative process feedback parameter.  

A convenient way to monitor the quality of an emulsion is by calculating D values, which are related to the cumulative distribution function (CDF) of a PSD and typically followed by a number. Consider for example D10, D50, and D90, which indicate the diameter below which 10\%, 50\%, and 90\% of the entire particle volume are measured. More specifically, a D50 value of 500 nm means that 50\% of the measured volume is composed of particles smaller than 500 nm. Once the CDF is obtained, one can extract the particle size from the corresponding percentiles. An example is shown in Fig.~\ref{fig:emulsionmonitoring}(c) where the CDF is plotted (solid line) to the corresponding volume distribution shown in (d). The calculated D values are marked on the horizontal axes (crosses). When comparing the D values in panels (d) and (e), respectively, we observe a significant shift towards larger particle diameters for all three parameters. Finally, in Fig.~\ref{fig:emulsionmonitoring}(b) we report D values as a function of time, which are derived from the CDFs, each averaged over a time period of $t_{\rm{meas}} = 60~\rm s$. We first observe a relatively constant behavior of the D values up until $t_{\rm{cp}}$, where we can see a significant increase in all three parameters. The time-resolved plot of the D values reveals interesting insights, in particular the D90 value, which is very sensitive to large particle diameters, as indicated by the fluctuations during the monitoring process.

To summarize this part, we have demonstrated the capability of \ofi for real-time monitoring of PSDs and D values. Moreover, our results prove that the system is capable of detecting changes in the dynamics of industrial relevant samples.

\subsection{Monitoring of Ni-P/SiC electrolyte solution}
We are currently implementing \ofi in collaboration with one of our industrial partners for continuous monitoring of SiC nanoparticles in an electrolyte solution used for coatings within electroplating processes. Before installing our system into the pilot plant, we performed measurements with a down scaled version of the process at laboratory scale. It is known from the industrial partner that the nanoparticles start to agglomerate during the process, which has an impact on the final coating. One of our goals is to monitor and reduce the amount of agglomerates in response to changing process parameters, such as temperature and pH. For this purpose, we have prepared 2~L of Ni-P/SiC electrolyte solution (see Sec.~\ref{sec:prepsic} for details). Before measurement, we continuously diluted the sample with an in-house developed dilution system (see Sec.~\ref{sec:sampledilution}). In this experiment, the flow rate was set to $\dot{V} = 50~\mu\rm{L/min}$ for the measurement of SiC nanoparticles. 

Fig.~\ref{fig:SiCmonitoring} shows results for the trajectories of (a,c) PS and (b,d) SiC nanoparticles, obtained by \ofi. Usually, when performing measurements with \ofi, one obtains  waterfall diagrams from the light scattering of particles (see also Sec.~\ref{sec:dataprocessing}), as shown in panel (a) on the example of the PS nanospheres. From the diagram we can clearly identify single particles entering and leaving the observation window. Things somewhat change for the case of SiC nanoparticles shown in panel (b), where we observe dissociation of particle agglomerates as they approach the focal region $z_f$ (gray dashed line) of the laser beam.  In this particular case, we can observe three such events at once. We attribute the dissociation to large optical forces acting on the agglomerates. As discussed above, we obtain the particle velocity and subsequently size from the slope of the single particle trajectories. In the case of particle dissociation, however, we can no longer reliably infer the particle size from the trajectories. Here we propose a different approach where we introduce a novel feedback parameter based on the number of particles detected in our system. Within a time frame of $t_{\rm{meas}} = 6~\rm s$, we count $N_0$ and $N_1$ particles, respectively, at reference points $z_0$ and $z_1$ (dashed lines). 

Figure~\ref{fig:SiCmonitoring}(c) shows the difference in particle count $N_1 - N_0$ (solid line) for the case of PS nanoparticles where we observe positive and negative values around zero. As apparent from the figure, we miss some of the particles at the boundaries of the observation window, however, on average this should be negligible. As expected, we observe for the PS nanoparticles a constant average around zero (dashed line). The results for the SiC nanoparticles are shown in (d) and reveal a mean value significantly above zero, as well as more frequent positive peaks in the staircase plot. This allows us to estimate the time point of agglomeration dissociation and from the peak height the amount of particles involved in this process.

When analyzing panel (b) in more detail, we see dissociation of particle agglomerates happening at different positions relative to the focal region $z_f$ of the exciting laser beam. From this information, we aim to extract information about the dissociation constants of particle agglomerates. This would require further experiments as well as some extensions to our theoretical framework to gain more information on how the optical forces act in particle agglomerates, which we leave to future work.

\section{Summary \& Outlook}
In summary, we have presented \ofi for the continuous monitoring and real-time process feedback of industrially relevant samples. \ofi builds on a microfluidic flow channel and a higher order laser beam that allows characterization of samples with a high degree of polydispersity and multi modal PSDs over a broad size range. We have demonstrated the applicability of our system on the example of PS nanospheres, oil-in-water emulsions, and SiC nanoparticles immersed in an electrolyte solution. Our results support \ofi's unique measurement capabilities by providing process feedback parameters such as particle count, size and size distributions, and concentration with single-particle sensitivity to reveal process dynamics in real-time. Exploiting the active principle of this approach, we introduced a novel feedback parameter based on dissociation processes of particle agglomerates. With some extensions to our current approach, this could provide us with detailed information about dissociation constants of particle agglomerates in different electrolyte compositions.

We are currently working on extracting information about the forward scattered light which is especially observed for the case of large particles. This might provide us with additional information about particle size and shape of ultra-low concentrated particles or large particle counts (LPC) encountered in the micrometer size range. In the future, we plan to extend our characterization process feedback to include correlative particle analytics with spectroscopic information of individual particles, where we have already achieved first promising results on the example of nanoplastics. With these extensions, \ofi will provide an even more versatile workbench for a wide range of applications in both research and industry.

\subsection*{Acknowledgements}
We are grateful to Doris Auer, Alexander Leljak, Nikola \v{S}imi{\'c}, Michael Peinhopf, Michael Schnur, and Gerhard Prossliner for the excellent technical input within the BRAVE Analytics team.  We thank the whole nano-medicine workgroup at the Gottfried Schatz Research Center and Creative Nano PC for their cooperation and helpful discussions. 

\section*{Declarations}
\begin{description}
	\item[Funding] This work was supported in part by the Austrian Research Promotion Agency (FFG) through the projects AoDiSys 891714 and Nano-VISION 895429, the European Commission (EC) through the projects NanoPAT (H2020-NMBP-TO-IND-2018-2020, Grant Agreement number: 862583) and MOZART (HORIZON-CL4-2021-RESILIENCE-01, Grant Agreement Number: 101058450).
	\item[Author contributions] All authors contributed to the study conception and design. Material preparation and data collection was performed by Christian Neuper. Analysis, theory, and writing of the first draft of the manuscript was performed by Marko \v{S}imi{\'c}. All authors commented on previous versions, read, and approved the final manuscript.
	\item[Conflict of Interest] Marko \v{S}imi{\'c}, Christian Neuper and Christian Hill are affiliated with BRAVE Analytics GmbH, the exclusive licensing partner of the OF2i patent portfolio and supplier of OF2i instruments. Christian Hill is shareholder of BRAVE Analytics GmbH. Ulrich Hohenester declares no conflicts of interest.
	\item[Availability of data and materials] The datasets generated during and/or analysed during the current study are available from the corresponding author on reasonable request.
	\item[Code availability] The codes used for analysis are available from the corresponding author on reasonable request.
	\item[Consent to participate] Not applicable.
	\item[Consent for publication] Not applicable.
	\item[Ethics approval] Not applicable.
	
\end{description}


\begin{thebibliography}{24}
	\expandafter\ifx\csname natexlab\endcsname\relax\def\natexlab#1{#1}\fi
	\expandafter\ifx\csname bibnamefont\endcsname\relax
	\def\bibnamefont#1{#1}\fi
	\expandafter\ifx\csname bibfnamefont\endcsname\relax
	\def\bibfnamefont#1{#1}\fi
	\expandafter\ifx\csname citenamefont\endcsname\relax
	\def\citenamefont#1{#1}\fi
	\expandafter\ifx\csname url\endcsname\relax
	\def\url#1{\texttt{#1}}\fi
	\expandafter\ifx\csname urlprefix\endcsname\relax\def\urlprefix{URL }\fi
	\providecommand{\bibinfo}[2]{#2}
	\providecommand{\eprint}[2][]{\url{#2}}
	
	\bibitem[{\citenamefont{Mitchell et~al.}(2020)\citenamefont{Mitchell,
			Billingsley, Haley, Wechsler, Peppas, and Langer}}]{mitchell:20}
	\bibinfo{author}{\bibfnamefont{M.~J.} \bibnamefont{Mitchell}},
	\bibinfo{author}{\bibfnamefont{M.~M.} \bibnamefont{Billingsley}},
	\bibinfo{author}{\bibfnamefont{R.~M.} \bibnamefont{Haley}},
	\bibinfo{author}{\bibfnamefont{M.~E.} \bibnamefont{Wechsler}},
	\bibinfo{author}{\bibfnamefont{N.~A.} \bibnamefont{Peppas}},
	\bibnamefont{and} \bibinfo{author}{\bibfnamefont{R.}~\bibnamefont{Langer}},
	\bibinfo{journal}{Nature Reviews Drug Discovery}
	\textbf{\bibinfo{volume}{20}}, \bibinfo{pages}{101} (\bibinfo{year}{2020}).
	
	\bibitem[{\citenamefont{Gupta et~al.}(2022)\citenamefont{Gupta, Mohapatra,
			Mishra, Farooq, Kumar, Ansari, Aldawsari, Alalaiwe, Mirza, and
			Iqbal}}]{gupta:22}
	\bibinfo{author}{\bibfnamefont{V.}~\bibnamefont{Gupta}},
	\bibinfo{author}{\bibfnamefont{S.}~\bibnamefont{Mohapatra}},
	\bibinfo{author}{\bibfnamefont{H.}~\bibnamefont{Mishra}},
	\bibinfo{author}{\bibfnamefont{U.}~\bibnamefont{Farooq}},
	\bibinfo{author}{\bibfnamefont{K.}~\bibnamefont{Kumar}},
	\bibinfo{author}{\bibfnamefont{M.~J.} \bibnamefont{Ansari}},
	\bibinfo{author}{\bibfnamefont{M.~F.} \bibnamefont{Aldawsari}},
	\bibinfo{author}{\bibfnamefont{A.~S.} \bibnamefont{Alalaiwe}},
	\bibinfo{author}{\bibfnamefont{M.~A.} \bibnamefont{Mirza}}, \bibnamefont{and}
	\bibinfo{author}{\bibfnamefont{Z.}~\bibnamefont{Iqbal}},
	\bibinfo{journal}{Gels} \textbf{\bibinfo{volume}{8}} (\bibinfo{year}{2022}),
	ISSN \bibinfo{issn}{2310-2861},
	\urlprefix\url{https://www.mdpi.com/2310-2861/8/3/173}.
	
	\bibitem[{\citenamefont{Kaiser et~al.}(2013)\citenamefont{Kaiser, Zuin, and
			Wick}}]{kaiser:13}
	\bibinfo{author}{\bibfnamefont{J.-P.} \bibnamefont{Kaiser}},
	\bibinfo{author}{\bibfnamefont{S.}~\bibnamefont{Zuin}}, \bibnamefont{and}
	\bibinfo{author}{\bibfnamefont{P.}~\bibnamefont{Wick}},
	\bibinfo{journal}{Science of The Total Environment}
	\textbf{\bibinfo{volume}{442}}, \bibinfo{pages}{282} (\bibinfo{year}{2013}),
	ISSN \bibinfo{issn}{0048-9697},
	\urlprefix\url{https://www.sciencedirect.com/science/article/pii/S0048969712012910}.
	
	\bibitem[{\citenamefont{Nile et~al.}(2020)\citenamefont{Nile, Baskar, Selvaraj,
			Nile, Xiao, and Kai}}]{nile2020nanotechnologies}
	\bibinfo{author}{\bibfnamefont{S.~H.} \bibnamefont{Nile}},
	\bibinfo{author}{\bibfnamefont{V.}~\bibnamefont{Baskar}},
	\bibinfo{author}{\bibfnamefont{D.}~\bibnamefont{Selvaraj}},
	\bibinfo{author}{\bibfnamefont{A.}~\bibnamefont{Nile}},
	\bibinfo{author}{\bibfnamefont{J.}~\bibnamefont{Xiao}}, \bibnamefont{and}
	\bibinfo{author}{\bibfnamefont{G.}~\bibnamefont{Kai}},
	\bibinfo{journal}{Nano-micro letters} \textbf{\bibinfo{volume}{12}},
	\bibinfo{pages}{1} (\bibinfo{year}{2020}).
	
	\bibitem[{\citenamefont{Finsy}(1994)}]{finsy:94}
	\bibinfo{author}{\bibfnamefont{R.}~\bibnamefont{Finsy}},
	\bibinfo{journal}{Advances in Colloid and Interface Science}
	\textbf{\bibinfo{volume}{52}}, \bibinfo{pages}{79} (\bibinfo{year}{1994}).
	
	\bibitem[{\citenamefont{Stetefeld et~al.}(2016)\citenamefont{Stetefeld,
			McKenna, and Patel}}]{stetefeld:16}
	\bibinfo{author}{\bibfnamefont{J.}~\bibnamefont{Stetefeld}},
	\bibinfo{author}{\bibfnamefont{S.~A.} \bibnamefont{McKenna}},
	\bibnamefont{and} \bibinfo{author}{\bibfnamefont{T.~R.} \bibnamefont{Patel}},
	\bibinfo{journal}{Biophys. Rev.} \textbf{\bibinfo{volume}{8}},
	\bibinfo{pages}{409} (\bibinfo{year}{2016}).
	
	\bibitem[{\citenamefont{Hole et~al.}(2013)\citenamefont{Hole, Sillence,
			Hannell, Maguire, Roesslein, Suarez, Capracotta, Magdolenova, Horev-Azaria,
			Dybowska et~al.}}]{hole:13}
	\bibinfo{author}{\bibfnamefont{P.}~\bibnamefont{Hole}},
	\bibinfo{author}{\bibfnamefont{K.}~\bibnamefont{Sillence}},
	\bibinfo{author}{\bibfnamefont{C.}~\bibnamefont{Hannell}},
	\bibinfo{author}{\bibfnamefont{C.~M.} \bibnamefont{Maguire}},
	\bibinfo{author}{\bibfnamefont{M.}~\bibnamefont{Roesslein}},
	\bibinfo{author}{\bibfnamefont{G.}~\bibnamefont{Suarez}},
	\bibinfo{author}{\bibfnamefont{S.}~\bibnamefont{Capracotta}},
	\bibinfo{author}{\bibfnamefont{Z.}~\bibnamefont{Magdolenova}},
	\bibinfo{author}{\bibfnamefont{L.}~\bibnamefont{Horev-Azaria}},
	\bibinfo{author}{\bibfnamefont{A.}~\bibnamefont{Dybowska}},
	\bibnamefont{et~al.}, \bibinfo{journal}{Journal of Nanoparticle Research}
	\textbf{\bibinfo{volume}{15}} (\bibinfo{year}{2013}).
	
	\bibitem[{\citenamefont{Anderson et~al.}(2013)\citenamefont{Anderson, Kozak,
			Coleman, J\"amting, and Trau}}]{anderson:13}
	\bibinfo{author}{\bibfnamefont{W.}~\bibnamefont{Anderson}},
	\bibinfo{author}{\bibfnamefont{D.}~\bibnamefont{Kozak}},
	\bibinfo{author}{\bibfnamefont{V.~A.} \bibnamefont{Coleman}},
	\bibinfo{author}{\bibfnamefont{{\AA}.~K.} \bibnamefont{J\"amting}},
	\bibnamefont{and} \bibinfo{author}{\bibfnamefont{M.}~\bibnamefont{Trau}},
	\bibinfo{journal}{Journal of Colloid and Interface Science}
	\textbf{\bibinfo{volume}{405}}, \bibinfo{pages}{322} (\bibinfo{year}{2013}).
	
	\bibitem[{iso()}]{iso:16}
	\bibinfo{note}{ISO/TR 18196:2016, Nanotechnologies -- Measurement technique
		matrix for the characterization of nano-objects (2016).}
	
	\bibitem[{\citenamefont{Jon{\'a}{\v{s}} and Zemanek}(2008)}]{jonavs2008light}
	\bibinfo{author}{\bibfnamefont{A.}~\bibnamefont{Jon{\'a}{\v{s}}}}
	\bibnamefont{and} \bibinfo{author}{\bibfnamefont{P.}~\bibnamefont{Zemanek}},
	\bibinfo{journal}{Electrophoresis} \textbf{\bibinfo{volume}{29}},
	\bibinfo{pages}{4813} (\bibinfo{year}{2008}).
	
	\bibitem[{\citenamefont{Ashkin}(1970)}]{ashkin:70}
	\bibinfo{author}{\bibfnamefont{A.}~\bibnamefont{Ashkin}},
	\bibinfo{journal}{Phys. Rev. Lett.} \textbf{\bibinfo{volume}{24}},
	\bibinfo{pages}{156} (\bibinfo{year}{1970}).
	
	\bibitem[{\citenamefont{B{\=u}tait{\.e}
			et~al.}(2019)\citenamefont{B{\=u}tait{\.e}, Gibson, Ho, Taverne, Taylor, and
			Phillips}}]{butaite2019indirect:19}
	\bibinfo{author}{\bibfnamefont{U.~G.} \bibnamefont{B{\=u}tait{\.e}}},
	\bibinfo{author}{\bibfnamefont{G.~M.} \bibnamefont{Gibson}},
	\bibinfo{author}{\bibfnamefont{Y.-L.~D.} \bibnamefont{Ho}},
	\bibinfo{author}{\bibfnamefont{M.}~\bibnamefont{Taverne}},
	\bibinfo{author}{\bibfnamefont{J.~M.} \bibnamefont{Taylor}},
	\bibnamefont{and} \bibinfo{author}{\bibfnamefont{D.~B.}
		\bibnamefont{Phillips}}, \bibinfo{journal}{Nature communications}
	\textbf{\bibinfo{volume}{10}}, \bibinfo{pages}{1} (\bibinfo{year}{2019}).
	
	\bibitem[{\citenamefont{Donato et~al.}(2016)\citenamefont{Donato, Mazzulla,
			Pagliusi, Magazz{\`u}, Hernandez, Provenzano, Gucciardi, Marag{\`o}, and
			Cipparrone}}]{donato2016light:16}
	\bibinfo{author}{\bibfnamefont{M.}~\bibnamefont{Donato}},
	\bibinfo{author}{\bibfnamefont{A.}~\bibnamefont{Mazzulla}},
	\bibinfo{author}{\bibfnamefont{P.}~\bibnamefont{Pagliusi}},
	\bibinfo{author}{\bibfnamefont{A.}~\bibnamefont{Magazz{\`u}}},
	\bibinfo{author}{\bibfnamefont{R.}~\bibnamefont{Hernandez}},
	\bibinfo{author}{\bibfnamefont{C.}~\bibnamefont{Provenzano}},
	\bibinfo{author}{\bibfnamefont{P.}~\bibnamefont{Gucciardi}},
	\bibinfo{author}{\bibfnamefont{O.}~\bibnamefont{Marag{\`o}}},
	\bibnamefont{and}
	\bibinfo{author}{\bibfnamefont{G.}~\bibnamefont{Cipparrone}},
	\bibinfo{journal}{Scientific reports} \textbf{\bibinfo{volume}{6}},
	\bibinfo{pages}{1} (\bibinfo{year}{2016}).
	
	\bibitem[{\citenamefont{Bustamante et~al.}(2021)\citenamefont{Bustamante,
			Chemla, Liu, and Wang}}]{bustamante2021optical}
	\bibinfo{author}{\bibfnamefont{C.~J.} \bibnamefont{Bustamante}},
	\bibinfo{author}{\bibfnamefont{Y.~R.} \bibnamefont{Chemla}},
	\bibinfo{author}{\bibfnamefont{S.}~\bibnamefont{Liu}}, \bibnamefont{and}
	\bibinfo{author}{\bibfnamefont{M.~D.} \bibnamefont{Wang}},
	\bibinfo{journal}{Nature Reviews Methods Primers}
	\textbf{\bibinfo{volume}{1}}, \bibinfo{pages}{25} (\bibinfo{year}{2021}).
	
	\bibitem[{\citenamefont{Imasaka et~al.}(1995)\citenamefont{Imasaka, Kawabata,
			Kaneta, and Ishidzu}}]{imasaka:95}
	\bibinfo{author}{\bibfnamefont{T.}~\bibnamefont{Imasaka}},
	\bibinfo{author}{\bibfnamefont{Y.}~\bibnamefont{Kawabata}},
	\bibinfo{author}{\bibfnamefont{T.}~\bibnamefont{Kaneta}}, \bibnamefont{and}
	\bibinfo{author}{\bibfnamefont{Y.}~\bibnamefont{Ishidzu}},
	\bibinfo{journal}{Analytical Chemistry} \textbf{\bibinfo{volume}{67}},
	\bibinfo{pages}{1763} (\bibinfo{year}{1995}).
	
	\bibitem[{\citenamefont{{\v{S}}imi{\'c}
			et~al.}(2022)\citenamefont{{\v{S}}imi{\'c}, Auer, Neuper, {\v{S}}imi{\'c},
			Prossliner, Prassl, Hill, and Hohenester}}]{simic:22}
	\bibinfo{author}{\bibfnamefont{M.}~\bibnamefont{{\v{S}}imi{\'c}}},
	\bibinfo{author}{\bibfnamefont{D.}~\bibnamefont{Auer}},
	\bibinfo{author}{\bibfnamefont{C.}~\bibnamefont{Neuper}},
	\bibinfo{author}{\bibfnamefont{N.}~\bibnamefont{{\v{S}}imi{\'c}}},
	\bibinfo{author}{\bibfnamefont{G.}~\bibnamefont{Prossliner}},
	\bibinfo{author}{\bibfnamefont{R.}~\bibnamefont{Prassl}},
	\bibinfo{author}{\bibfnamefont{C.}~\bibnamefont{Hill}}, \bibnamefont{and}
	\bibinfo{author}{\bibfnamefont{U.}~\bibnamefont{Hohenester}},
	\bibinfo{journal}{Physical Review Applied} \textbf{\bibinfo{volume}{18}},
	\bibinfo{pages}{024056} (\bibinfo{year}{2022}).
	
	\bibitem[{\citenamefont{Meinert et~al.}(2017)\citenamefont{Meinert, Gutwein,
			and Rohrbach}}]{meinert:17}
	\bibinfo{author}{\bibfnamefont{T.}~\bibnamefont{Meinert}},
	\bibinfo{author}{\bibfnamefont{B.~A.} \bibnamefont{Gutwein}},
	\bibnamefont{and} \bibinfo{author}{\bibfnamefont{A.}~\bibnamefont{Rohrbach}},
	\bibinfo{journal}{Opt. Lett.} \textbf{\bibinfo{volume}{42}},
	\bibinfo{pages}{350} (\bibinfo{year}{2017}).
	
	\bibitem[{\citenamefont{{\v{S}}imi{\'c}
			et~al.}(2023)\citenamefont{{\v{S}}imi{\'c}, Hill, and Hohenester}}]{simic:23}
	\bibinfo{author}{\bibfnamefont{M.}~\bibnamefont{{\v{S}}imi{\'c}}},
	\bibinfo{author}{\bibfnamefont{C.}~\bibnamefont{Hill}}, \bibnamefont{and}
	\bibinfo{author}{\bibfnamefont{U.}~\bibnamefont{Hohenester}},
	\bibinfo{journal}{Physical Review Applied} \textbf{\bibinfo{volume}{19}},
	\bibinfo{pages}{034041} (\bibinfo{year}{2023}).
	
	\bibitem[{\citenamefont{Kubo et~al.}(1985)\citenamefont{Kubo, Toda, and
			Hashitsume}}]{kubo:85}
	\bibinfo{author}{\bibfnamefont{R.}~\bibnamefont{Kubo}},
	\bibinfo{author}{\bibfnamefont{M.}~\bibnamefont{Toda}}, \bibnamefont{and}
	\bibinfo{author}{\bibfnamefont{M.}~\bibnamefont{Hashitsume}},
	\emph{\bibinfo{title}{Statistical Physics {II}}}
	(\bibinfo{publisher}{Springer}, \bibinfo{address}{Berlin},
	\bibinfo{year}{1985}).
	
	\bibitem[{\citenamefont{Einstein}(1905)}]{einstein:05}
	\bibinfo{author}{\bibfnamefont{A.}~\bibnamefont{Einstein}},
	\bibinfo{journal}{Annalen der Physik} \textbf{\bibinfo{volume}{322}},
	\bibinfo{pages}{549–560} (\bibinfo{year}{1905}).
	
	\bibitem[{\citenamefont{Neuman and Nagy}(2008)}]{neuman:08}
	\bibinfo{author}{\bibfnamefont{K.~C.} \bibnamefont{Neuman}} \bibnamefont{and}
	\bibinfo{author}{\bibfnamefont{A.}~\bibnamefont{Nagy}},
	\bibinfo{journal}{Nat. Methods} \textbf{\bibinfo{volume}{5}},
	\bibinfo{pages}{491} (\bibinfo{year}{2008}).
	
	\bibitem[{\citenamefont{Zhao et~al.}(2019)\citenamefont{Zhao, Wulder, Hu,
			Bright, Wu, Qin, Li, Toman, Mallick, Zhang et~al.}}]{zhao2019detecting}
	\bibinfo{author}{\bibfnamefont{K.}~\bibnamefont{Zhao}},
	\bibinfo{author}{\bibfnamefont{M.~A.} \bibnamefont{Wulder}},
	\bibinfo{author}{\bibfnamefont{T.}~\bibnamefont{Hu}},
	\bibinfo{author}{\bibfnamefont{R.}~\bibnamefont{Bright}},
	\bibinfo{author}{\bibfnamefont{Q.}~\bibnamefont{Wu}},
	\bibinfo{author}{\bibfnamefont{H.}~\bibnamefont{Qin}},
	\bibinfo{author}{\bibfnamefont{Y.}~\bibnamefont{Li}},
	\bibinfo{author}{\bibfnamefont{E.}~\bibnamefont{Toman}},
	\bibinfo{author}{\bibfnamefont{B.}~\bibnamefont{Mallick}},
	\bibinfo{author}{\bibfnamefont{X.}~\bibnamefont{Zhang}},
	\bibnamefont{et~al.}, \bibinfo{journal}{Remote sensing of Environment}
	\textbf{\bibinfo{volume}{232}}, \bibinfo{pages}{111181}
	(\bibinfo{year}{2019}).
	
	\bibitem[{\citenamefont{Sultanova et~al.}(2009)\citenamefont{Sultanova,
			Kasarova, and Nikolov}}]{sultanova:09}
	\bibinfo{author}{\bibfnamefont{N.}~\bibnamefont{Sultanova}},
	\bibinfo{author}{\bibfnamefont{S.}~\bibnamefont{Kasarova}}, \bibnamefont{and}
	\bibinfo{author}{\bibfnamefont{I.}~\bibnamefont{Nikolov}},
	\bibinfo{journal}{Acta Physica Polonica A} \textbf{\bibinfo{volume}{116}},
	\bibinfo{pages}{585} (\bibinfo{year}{2009}).
	
	\bibitem[{\citenamefont{Burgess et~al.}(2004)\citenamefont{Burgess, Duffy,
			Etzler, and Hickey}}]{burgess2004particle}
	\bibinfo{author}{\bibfnamefont{D.~J.} \bibnamefont{Burgess}},
	\bibinfo{author}{\bibfnamefont{E.}~\bibnamefont{Duffy}},
	\bibinfo{author}{\bibfnamefont{F.}~\bibnamefont{Etzler}}, \bibnamefont{and}
	\bibinfo{author}{\bibfnamefont{A.~J.} \bibnamefont{Hickey}},
	\bibinfo{journal}{The AAPS journal} \textbf{\bibinfo{volume}{6}},
	\bibinfo{pages}{23} (\bibinfo{year}{2004}).
	
\end{thebibliography}

\end{document}